\documentclass[journal]{IEEEtran}
\usepackage{url}
\usepackage{multirow}

\usepackage{cite}

\ifCLASSINFOpdf
  \usepackage[pdftex]{graphicx}
\else
  \usepackage[dvips]{graphicx}
\fi

\usepackage[cmex10]{amsmath}

\usepackage{stfloats}
\begin{document}
%
\title{Improved Parameter Estimation Techniques for Induction Motors using Hybrid Algorithms}
%
%
%

\author{Julius~Susanto,~\IEEEmembership{Member,~IEEE,}
        and~Syed~Islam,~\IEEEmembership{Senior~Member,~IEEE}
\thanks{Corresponding author: J. Susanto, e-mail: susanto@ieee.org}} 
\maketitle

\begin{abstract}
The performance of Newton-Raphson, Levenberg-Marquardt, Damped Newton-Raphson and genetic algorithms are investigated for the estimation of induction motor equivalent circuit parameters from commonly available manufacturer data. A new hybrid algorithm is then proposed that combines the advantages of both descent and natural optimisation algorithms. Through computer simulation, the hybrid algorithm is shown to significantly outperform the conventional algorithms in terms of convergence and squared error rates. All of the algorithms are tested on a large data set of 6,380 IEC (50Hz) and NEMA (60Hz) motors.

\end{abstract}


%
\IEEEpeerreviewmaketitle


\section{Introduction}
%
%
%
%
\IEEEPARstart{T}{he} three-phase induction motor is arguably the workhorse of modern industry, found in almost all industrial settings from manufacturing to mining. Equivalent circuit parameters of induction machines are essential for time-domain simulations where the dynamic interactions between the machine(s) and the power system need to be analysed, for example:

\begin{itemize}
\item Motor starting and re-acceleration
\item Bus transfer studies
\item Changes in motor loading
\item Motor behaviour during faults
\item Dynamic voltage stability
\end{itemize}

However, motor manufacturers do not tend to provide the equivalent circuit parameters for their machines. As the parameters are motor specific, typical values found in the literature are often not sufficiently accurate. Moreover, power system studies involving motors are normally performed during the design phases of projects, where the motors themselves have not yet been ordered and on-site testing is not possible. 

It is therefore desirable to estimate motor equivalent circuit parameters from the data that manufacturers make available in their catalogues, data sheets and technical brochures (e.g. performance parameters such as breakdown torque, locked rotor torque, full-load power factor, full-load efficiency, etc). 

A number of parameter estimation techniques have been proposed in the literature based on manufacturer data (for example \cite{johnson_1991}, \cite{rogers_1987}, \cite{waters_1983}, \cite{lindenmeyer_2001} and \cite{pedra_2004}). The de facto approach that has emerged, and which has been adopted by the majority of commercial software packages, has been to use an algorithm based on the Newton-Raphson method. However, it has been observed that the Newton-Raphson based algorithms can have poor convergence and error performance. Therefore, parameter estimation algorithms with improved performance are desired.

In this article, the performance of a number of parameter estimation algorithms based on readily available manufacturer data is investigated for the double-cage induction motor model. Section II describes the double-cage motor model and the formulation of the parameter estimation problem. In Section III, descent algorithms such as the Newton-Raphson method are discussed, followed by an overview of natural optimisation algorithms in Section IV. The properties of descent and natural optimisation algorithms are combined to overcome some of the limitations of descent algorithms and a new hybrid algorithm is proposed in Section V. The algorithms are then tested via computer simulation in Section VI and finally, the conclusions of this study are presented in Section VII.


\section{Parameter Estimation Problem for Induction Motors}
\subsection{Double Cage Induction Motor Model}
It has been previously shown that the double-cage model is appropriate for squirrel-cage induction motors in order to capture both the starting and breakdown performance of the motor \cite{vas_1992} \cite{pedra_2004}.

\begin{figure}[!t]
\centering
\includegraphics[width=2.5in]{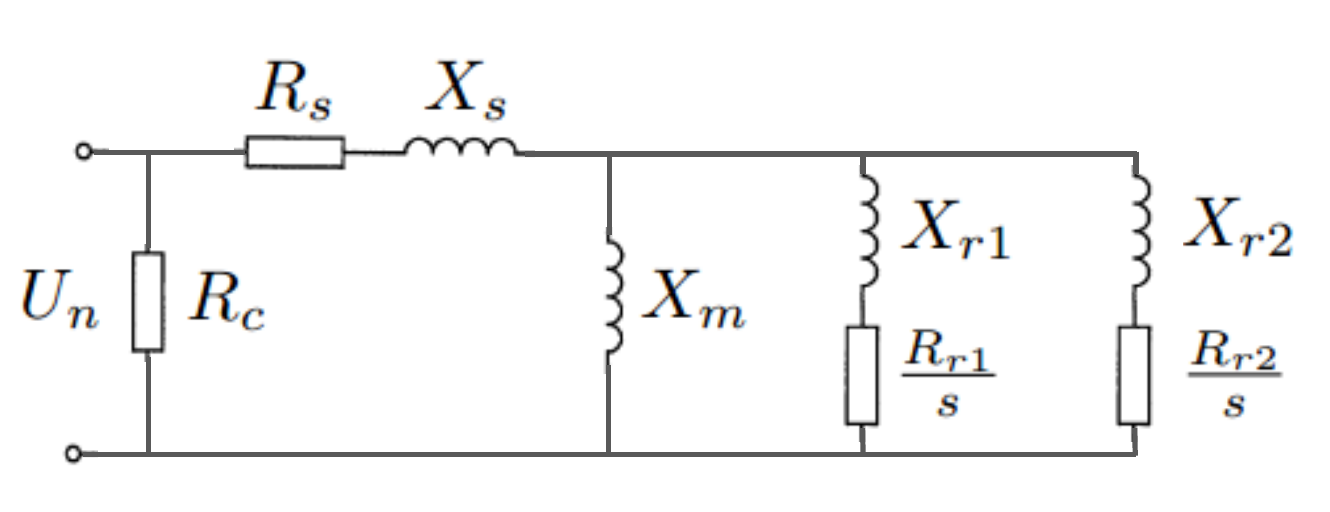}
\caption{Double cage equivalent circuit (eight parameter model)}
\label{fig:equiv_circuit}
\end{figure}

The double cage equivalent circuit model with eight slip-invariant parameters valid over the full range of slip values (i.e. from 0 to 1 pu) is shown in Figure \ref{fig:equiv_circuit}.

In the equivalent circuit, the inner cage leakage reactance $X_{r1}$ is always higher than the outer cage leakage reactance $X_{r2}$, but the outer cage impedance is typically higher than the inner cage impedance on starting. These conditions can be resolved by including the following two inequality constraints in the model \cite{pedra_2004}:

\begin{itemize}
\item $X_{r1} > X_{r2}$
\item $R_{r2} > R_{r1}$
\end{itemize}

In order to estimate motor efficiency, the core (and mechanical) losses also need to be included in the model. For simplicity, the core (and mechanical) losses are represented as a single shunt resistance $R_c$ at the input of the equivalent circuit \cite{pedra_2008}. 

\subsection{Parameter Estimation Problem Formulation}

The characteristics of an induction motor are normally provided by manufacturers in the form of standard performance parameters, with the following parameters being the most common:

\begin{itemize}
\item Nominal voltage, $U_n$ (V)
\item Nominal frequency, $f$ (Hz)
\item Rated asynchronous speed, $n_{fl}$ (rpm)
\item Rated (stator) current, $I_{s,fl}$ (A)
\item Rated mechanical power, $P_{m,fl}$ (kW)
\item Rated torque, $T_{n}$ (Nm)
\item Full load power factor, $\cos{\phi_{fl}}$ (pu)
\item Full load efficiency, $\eta_{fl}$ (pu)
\item Breakdown torque, $T_{b} / T_{n}$ (normalised)
\item Locked rotor torque, $T_{lr} / T_{n}$ (normalised)
\item Locked rotor current, $I_{lr} / I_{s,fl}$) (pu)
\end{itemize}

Given a set of performance parameters that contain features on the torque-speed and current-speed curves (e.g. breakdown torque, locked-rotor current, etc), the goal of a parameter estimation algorithm is to determine equivalent circuit parameters that yield these features.

While all of the performance parameters in the list above can be used in the estimation procedure, there are actually only six indpendent magnitudes that can be formed from them: $P_{m,fl}$, $Q_{fl}$, $T_{b}$, $T_{lr}$, $I_{lr}$ and $\eta_{fl}$ \cite{pedra_2004}. 

The six independent magnitudes can be used to formulate the parameter estimation problem in terms of a non-linear least squares problem, with a set of non-linear equations of the form $\mathbf{F}(\boldsymbol{x}) = \mathbf{0}$:

\begin{eqnarray}
f_{1} (\boldsymbol{x}) = P_{m,fl} - P(s_{f}) = 0 \\
f_{2} (\boldsymbol{x}) = Q_{fl} - Q(s_{f})  = 0 \\
f_{3} (\boldsymbol{x}) = T_{b} - T(s_{max}) = 0 \\
f_{4} (\boldsymbol{x}) = T_{lr} - T(s=1) = 0 \\
f_{5} (\boldsymbol{x}) = I_{lr} - I(s=1) = 0 \\
f_{6} (\boldsymbol{x}) = \eta{fl} - \eta(s_{f}) = 0
\end{eqnarray}

\noindent where $\mathbf{F} = ( f_1, f_2, f_3, f_4, f_5, f_6 )$ and \\ 
\indent $\boldsymbol{x} = ( R_s, X_s, X_m, R_{r1}, X_{r1}, R_{r2}, X_{r2}, R_{c} )$

In the formulation above, there are six independent equations in $\mathbf{F}$, but eight unknown parameters in $\boldsymbol{x}$. The non-linear system is therefore underdetermined, which leads to one of the key limitations in descent algorithms (refer to Section III for further discussion).

\subsection{Classes of Parameter Estimation Algorithms}
The parameter estimation problem formulated previously can be solved by a variety of non-linear least squares solver algorithms. As is usually the case with non-linear least squares problems, a closed form solution is not available and iterative algorithms are used to converge on a solution by minimising error residuals. 

Induction motor parameter estimation algorithms generally fall under two broad classes:
\begin{enumerate}
\item \textbf{Descent Methods}: are the class of algorithms based on variations of Newton's method for convergence to a solution, e.g. Newton-Raphson, Levenberg-Marquardt, etc
\item \textbf{Natural Optimisation Methods}: are the class of algorithms based on processes found in nature where successive randomised trials are filtered for "fitness" at each iteration, e.g. genetic algorithm, particle swarm optimisation, ant colony optimisation, simulated annealing, etc
\end{enumerate}

These classes of parameter estimation algorithms are discussed in more detail in the following sections.


\section{Descent Algorithms}
\subsection{Requirement for Linear Restrictions}
It was mentioned in Section II that the non-linear system of equations in the problem forumulation is underdetermined. Therefore, in order to make the systems exactly determined and solvable with descent algorithms, one must either:

\begin{enumerate}
\item Fix two parameters a priori (i.e. parameters are "known"), or;
\item Impose two constraints on the problem formulations, e.g. linear restrictions
\end{enumerate}

The use of linear restrictions was found to be superior to fixed parameters and therefore, all of the descent algorithms in this paper will include two linear restrictions by default. 

It was shown in \cite{pedra_2004} that the stator resistance $R_{s}$ was the least sensitive parameter in the equivalent circuit, i.e. variations in the value of $R_{s}$ had the least significant effect on the resulting torque-speed and current-speed curves. 
Therefore, $R_{s}$ can be subject to a linear restriction by linking it to the inner cage rotor resistance $R_{r1}$. Moreover, it is assumed that the outer cage rotor reactance $X_{r2}$ is linearly related to the stator reactance $X_{s}$. This leads to the following linear restrictions:

\begin{itemize}
\item $R_{s} = k_{r} R_{r1}$
\item $X_{r2} = k_{x} X_{s}$
\end{itemize}

\noindent Where $k_{r}$ and  $k_{x}$ are constant linear restrictions

\subsection{Newton-Raphson Algorithm}
\label{sec:nr_algo}
Of the class of descent methods used to solve non-linear least squares problems, the Newton-Raphson (NR) algorithm is probably the most straightforward. The NR algorithm is an iterative method where each iteration is calculated as follows:

\begin{equation}
\label{eqn_NR}
\boldsymbol{x}^{k+1} = \boldsymbol{x}^{k} - h_{n} \mathbf{J}^{-1}  \mathbf{F}( \boldsymbol{x}^{k})
\end{equation}

where $\boldsymbol{x}^{k+1}$ is the solution at the $(k+1)$th iteration, $\boldsymbol{x}^{k}$ is the solution at the $k$th iteration, $h_{n}$ is the step-size coefficient and $\mathbf{J}$ is the Jacobian matrix evaluated with the parameters at the $k$th iteration, $\boldsymbol{x}^{k}$.

The Jacobian matrix $\mathbf{J}$ has the general form:

\begin{equation}
\label{eqn_Jac}
\mathbf{J} = \left( \begin{array}{ccc}
\frac{\partial f_{1}}{\partial x_{1}}  & \ldots & \frac{\partial f_{1}}{\partial x_{6}} \\
\vdots & \ddots & \vdots \\
\frac{\partial f_{6}}{\partial x_{1}}  & \ldots & \frac{\partial f_{6}}{\partial x_{6}} \\ \end{array} \right)
\end{equation}

For systems where it is impractical to compute the exact partial derivatives analytically, a numerical approximation may be used with finite difference equations:

\begin{equation}
\frac{\partial f_{i}}{\partial x_{j}} \approx \frac{f_{i} (\mathbf{x} + \boldsymbol{\delta_{j}} h) - f_{i} (\mathbf{x})}{h}
\end{equation}

where $\boldsymbol{\delta_{j}}$ is vector of zeros with a single non-zero value of 1 at the j-th element and $h$ is a constant with a very small absolute value (e.g. $1 \times 10^{-6}$).

In this paper, a modified form of the NR algorithm proposed by Pedra in \cite{pedra_2008} for the double cage model is used as the default NR solver. This algorithm was selected because of its completeness, numerical accuracy and robustness compared to previously proposed methods (for example, in \cite{johnson_1991}, \cite{rogers_1987} and \cite{waters_1983}). Furthermore, the algorithm can be applied using commonly available manufacturer data, whereas other algorithms require more detailed data that may not be readily available (for example, multiple points on the torque-speed curve in \cite{lindenmeyer_2001}). 

\subsection{Levenberg-Marquardt Algorithm}
The Levenberg-Marquardt (LM) algorithm, sometimes called the damped least-squares algorithm, is another popular technique for solving least-squares problems \cite{levenberg_1944} \cite{marguardt_1963}. In the LM algorithm, each iteration is calculated as follows:

\begin{equation}
\label{eqn_LM}
\boldsymbol{x}^{k+1} = \boldsymbol{x}^{k} -  \left[ \mathbf{J}^{T} \mathbf{J} + \lambda \, \mathbf{diag}(\mathbf{J}^{T}\mathbf{J}) \right]^{-1} \mathbf{J}^{T} \mathbf{F}(\boldsymbol{x}^{k})
\end{equation}

where $\boldsymbol{x}^{k+1}$ is the solution at the $(k+1)$th iteration, $\boldsymbol{x}^{k}$ is the solution at the $k$th iteration, $\lambda$ is the damping parameter (more on this later) and $\mathbf{J}$ is the Jacobian matrix evaluated with the parameters at the $k$th iteration, $\boldsymbol{x}^{k}$ (as described previously in Equation \ref{eqn_Jac}).

\subsubsection{Choice of Damping Parameter}
The selection of the damping parameter $\lambda$ affects both the direction and magnitude of an iteration step. If the damping parameter is large, then the algorithm will move at short steps in the steepest descent direction. This is good when the iteration is far away from the solution. On the other hand, if the damping parameter is small, then the algorithm approaches a Gauss-Newton type method, which exhibits good convergence in the neighbourhood of the solution.

The damping parameter should thus be updated at each iteration depending on whether the algorithm is far or close to the solution. Two methods for adjusting the damping parameter are described below.

\paragraph{Gain Ratio Adjustment}
Marquardt suggested updating the damping parameter based on a "gain ratio" \cite{marguardt_1963}:

\begin{equation}
\rho = \frac{\mathbf{F}(\boldsymbol{x}^{k}) - \mathbf{F}(\boldsymbol{x}^{k+1})}{\frac{1}{2} \mathbf{\Delta x}^{T}(\lambda \mathbf{\Delta x} - \mathbf{J}^{T} \mathbf{F}(\boldsymbol{x}^{k}))}
\end{equation}

Where $\mathbf{\Delta x} = - \left[ \mathbf{J}^{T} \mathbf{J} + \lambda \, \mathbf{diag}(\mathbf{J}^{T}\mathbf{J}) \right]^{-1} \mathbf{J}^{T} \mathbf{F}(\boldsymbol{x}^{k})$ is the correction step at iteration k.

The damping parameter is adjusted depending on the value of the gain ratio as follows:

\begin{equation}
\lambda =
\begin{cases}
\lambda \times \beta, & \text{if } \rho < \rho_{1} \\
\frac{\lambda}{\gamma}, & \text{if } \rho > \rho_{2}
\end{cases}
\end{equation}

Where $\rho_{1}$, $\rho_{2}$, $\beta$ and $\gamma$ are algorithm control parameters. 

\paragraph{Error Term Adjustment}
An alternative to using the gain ratio is to adjust the damping parameter based only on the error term (i.e. the numerator of the gain ratio). The damping parameter is therefore updated as follows:

\begin{equation}
\lambda =
\begin{cases}
\lambda \times \beta, & \text{if } \mathbf{F}(\boldsymbol{x}^{k}) - \mathbf{F}(\boldsymbol{x}^{k+1}) < 0 \\
\frac{\lambda}{\gamma}, & \text{if } \mathbf{F}(\boldsymbol{x}^{k}) - \mathbf{F}(\boldsymbol{x}^{k+1}) > 0
\end{cases}
\end{equation}

Where $\rho_{1}$, $\rho_{2}$, $\beta$ and $\gamma$ are algorithm control parameters.

Computer simulations showed that the error term adjustment yielded significantly better performance than the gain ratio adjustment. Therefore in these studies, the error term adjustment is used exclusively along with the following algorithm control parameters: $\rho_{1} = 0.25$, $\rho_{2} = 0.75$, $\beta = 3$ and $\gamma = 3$.

\subsection{Damped Newton-Raphson Algorithm}
The damped Newton-Raphson algorithm is a variation of the conventional NR algorithm where a damping factor helps to get around problems with near-singular and/or ill-conditioned Jacobian matrices. In the damped NR algorithm, each iteration is calculated as follows:

\begin{equation}
\label{eqn_DNR}
\boldsymbol{x}^{k+1} = \boldsymbol{x}^{k} - h_{n} (\mathbf{J}^{-1} + \lambda I ) \mathbf{F}( \boldsymbol{x}^{k})
\end{equation}

Where the damping parameter $\lambda$ is adjusted at each iteration based on the error term as follows:

\begin{equation}
\lambda =
\begin{cases}
\lambda \times \beta, & \text{if } \mathbf{F}(\boldsymbol{x}^{k}) - \mathbf{F}(\boldsymbol{x}^{k+1}) < 0 \\
\frac{\lambda}{\gamma}, & \text{if } \mathbf{F}(\boldsymbol{x}^{k}) - \mathbf{F}(\boldsymbol{x}^{k+1}) > 0
\end{cases}
\end{equation}


\section{Natural Optimisation Algorithms}
\subsection{Genetic Algorithm}
In the context of motor parameter estimation, the genetic algorithm (GA) is used to minimise the squared error of the problem formulation vector $\mathbf{F}$. In GA terminology, the squared error is called the "fitness function" and is calculated as follows:

\begin{equation}
fitness = \mathbf{F} \mathbf{F}'
\end{equation}

\noindent where $\mathbf{F} = ( f_1, f_2, f_3, f_4, f_5, f_6 )$

Genetic algorithms can be binary coded where the solution paramaters are quantized into binary strings (for example, in \cite{nangsue_1999}, \cite{weatherford_2003} and \cite{nolan_1994}). However, the equivalent circuit parameters in a motor are continuous parameters and not naturally quantized. Thus, binary coding necessarily imposes limits on the precision of the parameters (i.e. due to the chosen length of the binary string). For this reason, a continuous parameter algorithm is used in this study. A description of the genetic algorithm as it applies to motor parameter estimation follows.

An initial population of $n_{pop}$ parameter estimates are randomly sampled from a uniform distribution with upper and lower limits as shown in Table \ref{tab:ga_range}.

\begin{table}[!t]
	\renewcommand{\arraystretch}{1.3}
	\caption{Range of initial parameter estimates}
	\label{tab:ga_range}
	\begin{center}
	\begin{tabular}{|c|c|c|}
        \hline
		  \multirow{2}{*}{Parameter} & \multicolumn{2}{|c|}{Range of Initial Estimate (pu)} \\ \cline{2-3}
        & Lower Bound & Upper Bound \\ \hline
        $R_{s}$  	& 0  & 0.15	\\ \hline
        $X_{s}$  	& 0  & 0.15	\\ \hline
        $X_{m}$  	& 0  & 5	\\ \hline
        $R_{r1}$  	& 0  & 0.15	\\ \hline
        $X_{r1}$  	& 0  & 0.30	\\ \hline
        $R_{r2}$  	& 0  & 0.15	\\ \hline
        $X_{r2}$  	& 0  & 0.15	\\ \hline
        $R_{c}$  	& 0  & 100	\\ \hline
    \end{tabular}
	\end{center}
\end{table}

The fitness of each member in the population is then calculated and ranked. The lowest fitness members are discarded and the rest are retained to form the mating pool for the next generation (there are $n_{pool}$ members in the mating pool).

The fittest $n_{e}$ members in the mating pool are retained for the next generation as \textbf{elite children}. 

Of the remaining $n_{pop} - n_{e}$ children to be created for the next generation, $c_{f}$\% will be produced by crossover and the rest ($1 - c_{f}$\%) by mutation. The proportion $c_{f}$ is called the \textbf{crossover fraction}.

\begin{enumerate}
\item \textbf{Crossover}: in the crossover process, two members of the mating pool are randomly selected and combined by taking a random blend of each member's parameters.

\item \textbf{Mutation}: in the mutation process, a member of the mating pool is randomly selected and its parameters are mutated by adding Gaussian noise with parameter-dependent standard deviations (see Table \ref{tab:ga_mutation}).
\end{enumerate}

\begin{table}[!t]
	\renewcommand{\arraystretch}{1.3}
	\caption{Standard deviations for mutation noise}
	\label{tab:ga_mutation}
	\begin{center}
	\begin{tabular}{|c|c|}
        \hline
		  Parameter & Standard Deviation ($\sigma$) \\ \hline
        $R_{s}$  	& 0.01  	\\ \hline
        $X_{s}$  	& 0.01  	\\ \hline
        $X_{m}$  	& 0.33 	\\ \hline
        $R_{r1}$  	& 0.01  \\ \hline
        $X_{r1}$  	& 0.01  	\\ \hline
        $R_{r2}$  	& 0.01  	\\ \hline
        $X_{r2}$  	& 0.01  	\\ \hline
        $R_{c}$  	& 6.67  	\\ \hline
    \end{tabular}
	\end{center}
\end{table}

The fitness of the next generation is then calculated and the process repeats itself for $n_{gen}$ generations.

The default settings for the genetic algorithm implemented in this project for motor parameter estimation are shown in Table \ref{tab:ga_default}.

\begin{table}[!t]
	\renewcommand{\arraystretch}{1.3}
	\caption{Default settings for genetic algorithm}
	\label{tab:ga_default}
	\begin{center}
	\begin{tabular}{|c|c|c|}
        \hline
        Setting  	& Setting Description  & Default Value	\\ \hline
        $n_{pop}$  	& Population of each generation  & 20	\\ \hline
        $n_{pool}$  	& Number of members in the mating pool  & 15 \\ \hline
        $n_{e}$  	& Number of elite children  & 2	\\ \hline
        $c_{f}$  	& Crossover fraction  & 80\%	\\ \hline
    \end{tabular}
	\end{center}
\end{table}

\subsection{Other Natural Optimisation Algorithms}
Since the popularisation of the genetic algorithm in the 1980s, a number of other natural optimisation algorithms have been introduced. All of these algorithms are inspired by natural processes and all share the basic methodology of injecting randomness and selecting for fitness in order to iteratively reach an optimal point in the search-space.

The literature has numerous examples of natural optimisation algorithms adapted for the estimation of induction motor parameters, for example:

\begin{itemize}
\item Simulated annealing \cite{bhuvaneswari_2005}
\item Particle swarm optimisation \cite{sakthivel_2010a} \cite{sakthivel_2010c}
\item Artificial immune system \cite{sakthivel_2010}
\item Bacterial foraging technique \cite{sakthivel_2010b}
\item Ant colony optimisation \cite{chen_2008}
\item Harmony search \cite{marques_2012}
\end{itemize}

The results of these investigations indicate that while some of the alternative algorithms lead to faster convergence than the genetic algorithm, they tend to converge to the same solution and error rates. The aim of this study is to improve error rates and for this reason, alternative natural optimisation algorithms are not explored further.


\section{Proposed Hybrid Algorithm}
In the descent algorithms, linear restrictions are imposed on $R_s$ and $X_{r2}$ in order to make the underdetermined system of equations solvable. It was shown in \cite{corcoles_2002} that the double cage model with core losses has 8 minimum independent variables (MIVs). Therefore, by constraining $R_s$ and $X_{r2}$ with linear restrictions, the solution space is also constrained by two degrees of freedom. Therefore, a solution could potentially exist to an otherwise non-converging problem were the linear restrictions removed.

\begin{figure}[!t]
\begin{center}
\includegraphics[scale=0.45]{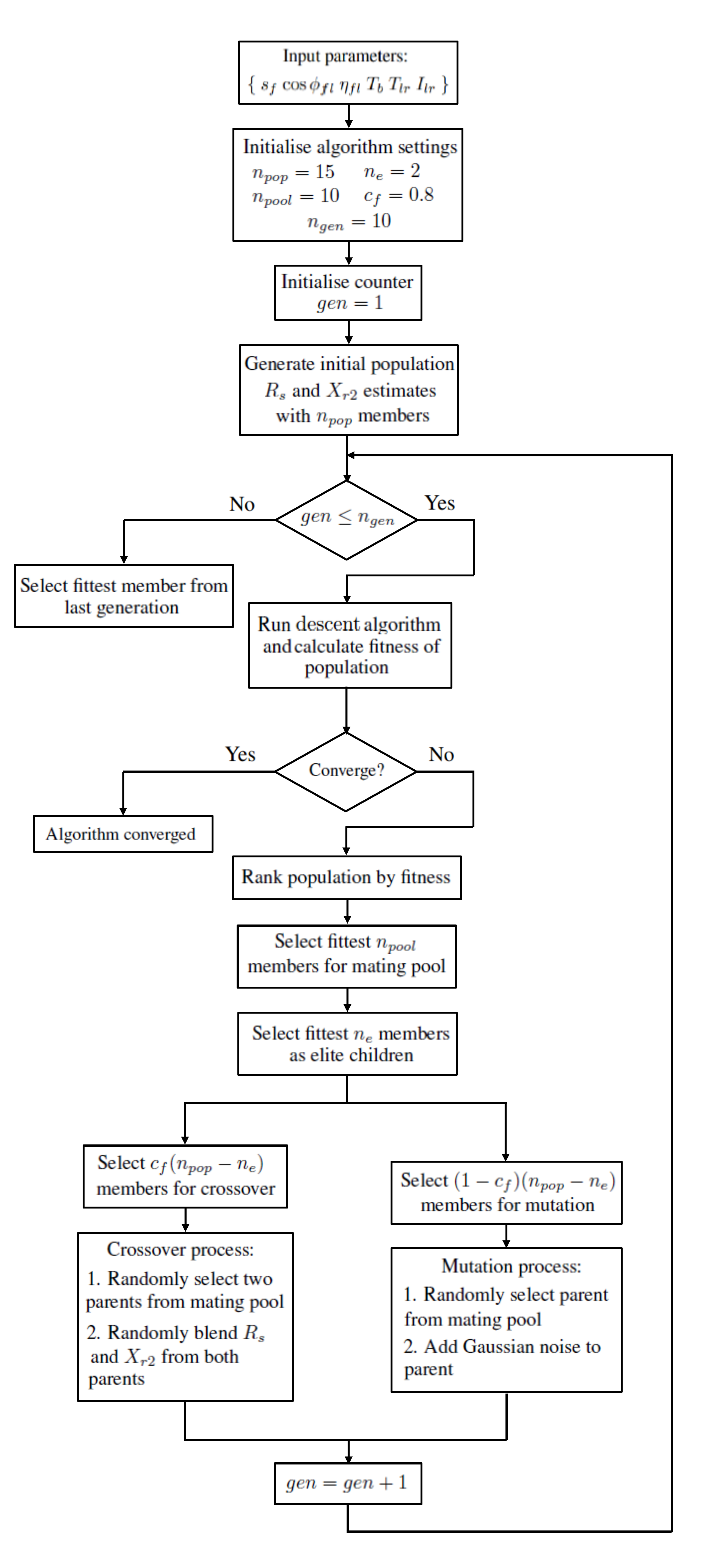}
\caption{Flowchart for hybrid algorithm (with natural selection of $R_{s}$ and $X_{r2}$)}
\label{fig:hybrid_algorithm}
\end{center}
\end{figure}

The proposed hybrid algorithm attempts to overcome this limitation of descent algorithms by applying a genetic algorithm to select $R_s$ and $X_{r2}$. In other words, a baseline descent algorithm (e.g. NR, Damped NR, LM, etc) is run with fixed values for $R_s$ and $X_{r2}$, which are in turn iteratively selected using a genetic algorithm in an outer loop. A flowchart of the proposed hybrid algorithm is shown in Figure \ref{fig:hybrid_algorithm}. A more detailed description of the proposed algorithm follows.

An initial population of $n_{pop}$ estimates for $R_s$ and $X_{r2}$ are randomly sampled from a uniform distribution with upper and lower limits as shown in Table \ref{tab:hybridga_range}. Each pair of $R_s$ and $X_{r2}$ estimates is referred to as a member of the population.

\begin{table}[!h]
	\renewcommand{\arraystretch}{1.3}
	\caption{Range of initial parameter estimates}
	\label{tab:hybridga_range}
	\begin{center}
	\begin{tabular}{|c|c|c|}
        \hline
		  \multirow{2}{*}{Parameter} & \multicolumn{2}{|c|}{Range of Initial Estimate (pu)} \\ \cline{2-3}
        & Lower Bound & Upper Bound \\ \hline
        $R_{s}$  	& 0  & 0.15	\\ \hline
        $X_{r2}$  	& 0  & 0.15	\\ \hline
    \end{tabular}
	\end{center}
\end{table}

The descent algorithm is then run on each member of the population. The fitness of each member (in terms of the squared error $\mathbf{F}'\mathbf{F}$) is calculated and ranked. As in the conventional genetic algorithm, the lowest fitness members are discarded and the rest are retained to form the mating pool for the next generation.

The fittest $n_{e}$ members in the mating pool are retained for the next generation as \textbf{elite children}. Of the remaining $n_{pop} - n_{e}$ children to be created for the next generation, $c_{f}$\% will be produced by crossover and the rest ($1 - c_{f}$\%) by mutation. The proportion $c_{f}$ is called the \textbf{crossover fraction}.

\begin{enumerate}
\item \textbf{Crossover}: in the crossover process, two members of the mating pool are randomly selected and combined by taking a random blend of each member's parameters, e.g. the crossover of parameter $R_{s}$:

\begin{equation}
	R_{s, child} = \alpha R_{s, parent 1} + (1 - \alpha) R_{s, parent 2}
\end{equation}

\noindent where $\alpha$ is a random variable selected from a uniform distribution over the interval $[0,1]$

\item \textbf{Mutation}: in the mutation process, a member of the mating pool is randomly selected and its parameters are mutated by adding Gaussian noise with standard deviations of 0.01.
\end{enumerate}

The descent algorithm is then run for the next generation of estimates for $R_s$ and $X_{r2}$. The fitness is calculated and the process repeats itself for $n_{gen}$ generations. If at any point during the process the descent algorithm converges, then the hybrid algorithm stops and selects the parameter estimates from the converged descent algorithm as the solution. Otherwise, the parameter estimates yielding the best fitness after $n_{gen}$ generations are selected.

\begin{table}[htp]
	\renewcommand{\arraystretch}{1.3}
	\caption{Default settings for proposed hybrid algorithms}
	\label{tab:hybrid_default}
	\begin{center}
	\begin{tabular}{|c|c|c|}
        \hline
        Setting  	& Setting Description  & Default Value	\\ \hline
        $n_{pop}$  	& Population of each generation  & 15	\\ \hline
        $n_{pool}$  	& Number of members in the mating pool  & 10 \\ \hline
        $n_{e}$  	& Number of elite children  & 2	\\ \hline
        $c_{f}$  	& Crossover fraction  & 80\%	\\ \hline
        $n_{gen}$  	& Maximum number of generations  & 10	\\ \hline
    \end{tabular}
	\end{center}
\end{table}

The following hybrid algorithms are investigated in this paper:

\begin{itemize}
\item NR-GA - Conventional NR and GA selection of $R_{s}$ and $X_{r2}$
\item LM-GA - Levenberg-Marquardt (with an error term lambda adjustment) and GA selection of $R_{s}$ and $X_{r2}$ 
\item DNR-GA - Damped NR (with a maximum number of 30 iterations for each damped NR step) and GA selection of $R_{s}$ and $X_{r2}$
\end{itemize}

The default settings applied for the proposed hybrid algorithms are shown in Table \ref{tab:hybrid_default}. 


\section{Computer Simulations}
\label{sec:comp_sim}

\subsection{Motor Data Set}
The algorithms presented in this paper were tested by computer simulation on a large data set from the EuroDEEM and MotorMaster databases (version 1.0.17 - 4 April 2007) \cite{eurodeem}. The original data set included over 15,000 IEC and NEMA type motors. From the original set, the motor data was conditioned by eliminating duplicate records, removing motors without power factor, efficiency or torque data and removing motors with strange or inconsistent data (e.g. full load torque greater than breakdown torque, asynchronous speed greater than synchronous speed, etc). After data cleansing, the final data set consisted of motors with nominal ratings from 0.37kW to 1000kW, and the following total quantities:

\begin{itemize}
\item 4,002 IEC 50Hz motors
\item 2,378 NEMA 60Hz motors
\end{itemize}

\subsection{Simulation Results}
A summary of the computer simulation results is shown in Table \ref{tab:results}. The table depicts the convergence rate, average squared error and maximum squared error for each of the algorithms described in this paper. Note that in this study, the convergence criterion is a squared error value of $1 \times 10^{-5}$ and the maximum number of iterations for the descent algorithms is 30.

\begin{table*}[!htp]
	\renewcommand{\arraystretch}{1.3}
	\caption{Simulation results for all algorithms}
	\label{tab:results}
	\begin{center}
	\begin{tabular}{|p{5cm}|p{2cm}|c|c|p{2cm}|c|c|}
        \hline
		  \multirow{3}{*}{Case} & \multicolumn{3}{|c|}{IEC Motors} & \multicolumn{3}{|c|}{NEMA Motors} \\ \cline{2-7}
        & \multirow{2}{*}{Convergence} & \multicolumn{2}{|c|}{Squared Error} & \multirow{2}{*}{Convergence} & \multicolumn{2}{|c|}{Squared Error} \\ \cline{3-4} \cline{6-7}
		  & & Average & Maximum & & Average & Maximum \\ \hline
        Newton-Raphson ($k_x = 1$,  $k_r = 0.5$) 	& 685 (17.1\%)  & 0.5411 & 5.9996	& 751 (31.6\%)	& 0.2514 & 5.9902  \\ \hline
		  Levenberg-Marquardt ($k_x = 1$,  $k_r = 0.5$) 	& 740 (18.5\%)  & 0.9114	& 6  & 770 (32.4\%)		& 0.2867 & 6  \\ \hline
		  Damped NR ($k_x = 1$,  $k_r = 0.5$) & 628 (15.7\%) & 0.2058 & 9.2314 & 568 (23.9\%)	& 0.0899	& 4.6188 \\ \hline
		  Newton-Raphson ($k_x = 0.5$,  $k_r = 1$) 	& 974 (24.3\%)  & 0.9261 & 120.08 & 934 (39.3\%)	& 0.1425 & 5.999 \\ \hline
		  Levenberg-Marquardt ($k_x = 0.5$,  $k_r = 1$) 	& 1035 (25.9\%) & 2.691 & 6 & 945 (39.7\%) & 1.7268 & 6	\\ \hline
		  Damped NR ($k_x = 0.5$,  $k_r = 1$) & 1006 (25.1\%) & 0.04 & 4.4584 & 935 (39.3\%) & 0.05054 & 4.6574	\\ \hline
		  Genetic Algorithm (Max gens = 30) & 0 (0.0\%) & 0.0471 & 3.6459 & 0 (0.0\%) & 0.04029 & 0.50916	\\ \hline
		  Genetic Algorithm (Max gens = 50) & 0 (0.0\%) & 0.0281 & 2.031 & 0 (0.0\%) & 0.03279 & 0.45363	\\ \hline
		  Genetic Algorithm (Max gens = 100) & 0 (0.0\%) & 0.01861 & 0.8062 & 0 (0.0\%) & 0.02676 & 0.37716	\\ \hline
		  Hybrid NR-GA & 1363 (34.1\%) & 0.0625 & 5.9998 & 1159 (48.7\%) & 0.0282 & 0.9682	\\ \hline
		  Hybrid LM-GA & 1388 (34.7\%) & 1.0941 & 6 & 1181 (49.7\%) & 0.53387 & 6	\\ \hline
		  Hybrid DNR-GA & 1373 (34.3\%) & 0.0174 & 5.9995 & 1168 (49.1\%) & 0.01948 & 0.42282	\\ \hline
    \end{tabular}
	\end{center}
\end{table*}

In terms of convergence and error rates, it can be seen from Table \ref{tab:results} that the hybrid algorithms are superior to the other algorithms. However, there is no hybrid algorithm that clearly stands out as the best option. The LM-GA has the highest number of converging solutions, but when it fails, it yields poor results (as evidenced by the high average error rate). The DNR-GA is more consistent in terms of low error rates, but the convergence rate is also lower. 

The genetic algorithms with maximum number of generations $\geq$30 also yield low average and maximum error rates, but never low enough to qualify for convergence (as defined by a squared error of $<1 \times 10^{-5}$). However, the performance of the genetic algorithm is unaffected by the choice of linear restrictions / initial conditions.

On the other hand, the results show that the descent algorithms are significantly affected by the choice of linear restrictions $k_r$ and $k_x$, and yield inferior convergence and error rates when compared to the proposed hybrid algorithms. 

The simulations suggest that the LM algorithm can lead to a higher convergence rate compared to the conventional NR algorithm, but at the cost of a higher average squared error. The LM algorithm works well in the neighbourhood of the solution, but does not perform very well at the early stages, particularly when the initial estimates are far from the solution. The LM algorithm can also produce spectacularly bad results when the Jacobian matrix is ill-conditioned or near-singular.

The damped NR algorithm is intended to help address the issue of ill-conditioned and near-singular Jacobian matrices. Adding a damping factor $\lambda I$ to the Jacobian matrix makes it more likely to be invertible. However, while the damped NR algorithm leads to lower average squared error, it also takes more iterations to converge. 

\subsection{Algorithm Computation Time}
\label{sec:computation_time}
Indicative computation times for the different algorithms in this project are shown in Table \ref{tab:solution_time}. The computation times were obtained from simulations performed on a 2.1GHz Intel dual core processor with 2GB RAM and are presented here primarily for comparison.

\begin{table}[htp]
	\renewcommand{\arraystretch}{1.3}
	\caption{Average algorithm solution time }
	\label{tab:solution_time}
	\begin{center}
	\begin{tabular}{|c|c|c|}
        \hline
		  \multirow{2}{*}{Algorithm} & \multicolumn{2}{|c|}{Solution Time (s)} \\ \cline{2-3}
       & Average  	& Maximum 	\\ \hline
       Newton-Raphson  	& 0.257 	& 0.742 	\\ \hline
       Levenberg-Marquardt 	& 0.162 	& 0.332 	\\ \hline
		 Damped NR 	& 0.241 	& 0.427 	\\ \hline
		 Genetic Algorithm (10 Gens) 	& 0.257 	& 0.328 	\\ \hline
		 Genetic Algorithm (30 Gens) 	& 0.778 	& 0.947 	\\ \hline
		 Genetic Algorithm (50 Gens) 	& 1.356 	& 1.758 	\\ \hline
		 Genetic Algorithm (100 Gens) 	& 2.753 	& 3.036 	\\ \hline
		 Hybrid NR-GA 	& 24.395 	& 53.290 	\\ \hline
		 Hybrid LM-GA 	& 14.289 	& 42.198 	\\ \hline
		 Hybrid DNR-GA 	& 29.050 	& 64.568 	\\ \hline
    \end{tabular}
	\end{center}
\end{table}

From Table \ref{tab:solution_time}, it can be seen that the hybrid algorithms have average solution times between 50 and 100 times slower than the conventional descent algorithms (i.e. NR, LM and DNR algorithms). This is because the evolutionary part of the hybrid algorithm must run the descent algorithm multiple times for each generation. For example, based on the default settings as shown in Table \ref{tab:hybrid_default}, the hybrid algorithm may have to perform up to $n_{pop} \times n_{gen} = 10 \times 15 = 150$ descent algorithms. This would occur in the worst case condition when the hybrid algorithm fails to converge.

The genetic algorithm has average solution times that are dependent on the maximum number of generations to be simulated. For a low number of generations (e.g. 10), the GA solution times are comparable to that of the descent algorithms. The solution times increase more or less linearly as the maximum number of generations is increased.

\subsection{Algorithm Performance and Motor Rated Power}

The performance of the algorithms are analysed with the data sets broken down by motor rated power. Table \ref{tab:breakdown_rating} presents the breakdown of the IEC and NEMA motor data sets, showing the quantity of motors for various nominal power ranges.

\begin{table}[htp]
	\renewcommand{\arraystretch}{1.3}
	\caption{Breakdown of motor data sets by motor rated power}
	\label{tab:breakdown_rating}
	\begin{center}
	\begin{tabular}{|c|c|c|}
        \hline
		  Motor Rating & No. IEC Motors & No. NEMA Motors \\ \hline
       0.37kW - 3.6kW  	& 1208 & 630  	\\ \hline
       4kW - 15kW  		& 963 & 598  	\\ \hline
       18.5kW - 75kW  	& 973 & 741  	\\ \hline
       90kW - 185kW  	& 477 & 284  	\\ \hline
       200kW - 630kW  	& 355 & 123  	\\ \hline
       $>$630kW			  	& 26 & 2  	\\ \hline
       TOTAL			  	& 4002 & 2378  	\\ \hline
    \end{tabular}
	\end{center}
\end{table}

\begin{table*}[htp]
	\renewcommand{\arraystretch}{1.3}
	\caption{Algorithm performance broken down by rated power (IEC motors)}
	\label{tab:results_by_rating_iec}
	\begin{center}
\begin{tabular}{|p{2.5cm}|p{1cm}|p{0.9cm}|p{1cm}|p{0.9cm}|p{1cm}|p{0.9cm}|p{1cm}|p{0.9cm}|p{1cm}|p{0.9cm}|p{1cm}|p{0.9cm}|}
        \hline
		  \multirow{2}{*}{Case} & \multicolumn{2}{|c|}{0.37 - 3.6kW} & \multicolumn{2}{|c|}{4 - 15kW} & \multicolumn{2}{|c|}{18.5 - 75kW} & \multicolumn{2}{|c|}{90 - 185kW} & \multicolumn{2}{|c|}{200 - 630kW} & \multicolumn{2}{|c|}{$>$630kW} \\ \cline{2-13} 
        & Conv-ergence & Average Error\textsuperscript{2} & Conv-ergence & Average Error\textsuperscript{2} & Conv-ergence & Average Error\textsuperscript{2} & Conv-ergence & Average Error\textsuperscript{2} & Conv-ergence & Average Error\textsuperscript{2} & Conv-ergence & Average Error\textsuperscript{2} \\ \hline
        Newton-Raphson ($k_x = 1$,  $k_r = 0.5$) 	& 4 (0.33\%)  & 0.8323 & 52 (0.33\%)	& 0.4891	& 305 (31.4\%) & 0.5880 & 166 (34.8\%) & 0.1788 & 137 (38.6\%) & 0.0887 & 21 (80.8\%) & 0.0092  \\ \hline
		  Levenberg-Marquardt ($k_x = 1$,  $k_r = 0.5$) & 14 (1.16\%)   &   1.9423   &   62 (6.44\%)   &   0.6743   &   336 (34.5\%)   &   0.1055   &   190 (39.8\%)   &   0.3229   &   122 (34.4\%)   &   0.9954   &   16 (61.5\%)   &   1.6178 \\ \hline
		  Damped NR ($k_x = 1$,  $k_r = 0.5$) & 4 (0.33\%)   &   0.5279   &   36 (3.7\%)   &   0.1132   &   262 (26.9\%)   &   0.0447   &   160 (33.5\%)   &   0.0402   &   143 (40.3\%)   &   0.0401   &   23 (88.5\%)   &   0.0047  \\ \hline
		  Genetic Algorithm (Max gens = 30) &0 (0.0\%)   &   0.0830   &   0 (0.0\%)   &   0.0196   &   0 (0.0\%)   &   0.0266   &   0 (0.0\%)   &   0.0469   &   0 (0.0\%)   &   0.0550   &   0 (0.0\%)   &   0.0569 \\ \hline
		  Genetic Algorithm (Max gens = 50) & 0 (0.0\%)   &   0.0446   &   0 (0.0\%)   &   0.0157   &   0 (0.0\%)   &   0.0180   &   0 (0.0\%)   &   0.0251   &   0 (0.0\%)   &   0.0362   &   0 (0.0\%)   &   0.0313 \\ \hline
		  Hybrid NR-GA & 48 (3.97\%)   &   0.1842   &   245 (25.4\%)   &   0.0219   &   559 (57.5\%)   &   0.0035   &   281 (58.9\%)   &   0.0032   &   207 (58.3\%)   &   0.0040   &   23 (88.5\%)   &   0.0014 \\ \hline
		  Hybrid LM-GA & 149 (12.3\%) & 2.147 & 274 (28.5\%) & 1.196 & 405 (41.6\%) & 0.5132 & 303 (63.5\%) & 0.1923 & 240 (67.6\%) & 0.1185 & 17 (65.4\%) & 0.0005 \\ \hline
		  Hybrid DNR-GA & 140 (11.6\%) & 0.037 & 265 (27.5\%) & 0.014 & 404 (41.5\%) & 0.0104 & 303 (63.5\%) & 0.0021 & 244 (68.7\%) & 0.0014 & 17 (65.4\%) & 0.0014  \\ \hline
    \end{tabular}
	\end{center}
\end{table*}

\begin{table*}[bhtp]
	\renewcommand{\arraystretch}{1.3}
	\caption{Algorithm performance broken down by rated power (NEMA motors)}
	\label{tab:results_by_rating_nema}
	\begin{center}
\begin{tabular}{|p{2.5cm}|p{1cm}|p{0.9cm}|p{1cm}|p{0.9cm}|p{1cm}|p{0.9cm}|p{1cm}|p{0.9cm}|p{1cm}|p{0.9cm}|p{1cm}|p{0.9cm}|}
        \hline
		  \multirow{2}{*}{Case} & \multicolumn{2}{|c|}{0.37 - 3.6kW} & \multicolumn{2}{|c|}{4 - 15kW} & \multicolumn{2}{|c|}{18.5 - 75kW} & \multicolumn{2}{|c|}{90 - 185kW} & \multicolumn{2}{|c|}{200 - 630kW} & \multicolumn{2}{|c|}{$>$630kW} \\ \cline{2-13} 
        & Conv-ergence & Average Error\textsuperscript{2} & Conv-ergence & Average Error\textsuperscript{2} & Conv-ergence & Average Error\textsuperscript{2} & Conv-ergence & Average Error\textsuperscript{2} & Conv-ergence & Average Error\textsuperscript{2} & Conv-ergence & Average Error\textsuperscript{2} \\ \hline
        Newton-Raphson ($k_x = 1$,  $k_r = 0.5$) 	& 9 (1.43\%)   &   0.3225   &   100 (16.7\%)   &   0.3254   &   415 (56.0\%)   &   0.1455   &   162 (57.0\%)   &   0.1052   &   65 (52.9\%)   &   0.0469   &   0 (0.0\%)   &   0.2338  \\ \hline
		  Levenberg-Marquardt ($k_x = 1$,  $k_r = 0.5$) & 83 (13.2\%)   &   0.2788   &   158 (26.4\%)   &   0.2833   &   395 (53.3\%)   &   0.3086   &   102 (35.9\%)   &   0.3288   &   31 (25.2\%)   &   0.1653   &   1 (50.0\%)   &   0.0765 \\ \hline
		  Damped NR ($k_x = 1$,  $k_r = 0.5$) &  58 (9.21\%)   &   0.0820   &   110 (18.4\%)   &   0.0751   &   297 (40.1\%)   &   0.0722   &   74 (26.1\%)   &   0.1683   &   28 (22.8\%)   &   0.0863   &   1 (50.0\%)   &   0.2711 \\ \hline
		  Genetic Algorithm (Max gens = 30) & 0 (0.0\%)   &   0.0366   &   0 (0.0\%)   &   0.0362   &   0 (0.0\%)   &   0.0449   &   0 (0.0\%)   &   0.0515   &   0 (0.0\%)   &   0.0493   &   0 (0.0\%)   &   0.0926 \\ \hline
		  Genetic Algorithm (Max gens = 50) & 0 (0.0\%)   &   0.0306   &   0 (0.0\%)   &   0.0303   &   0 (0.0\%)   &   0.0374   &   0 (0.0\%)   &   0.0405   &   0 (0.0\%)   &   0.0344   &   0 (0.0\%)   &   0.0872 \\ \hline
		  Hybrid NR-GA & 161 (25.6\%)   &   0.0317   &   311 (52.0\%)   &   0.0245   &   512 (69.1\%)   &   0.0226   &   123 (43.3\%)   &   0.0294   &   51 (41.5\%)   &   0.0059   &   1 (50.0\%)   &   0.0416 \\ \hline
		  Hybrid LM-GA & 237 (37.6\%) & 0.519 & 308 (51.5\%) & 0.5374 & 413 (55.7\%) & 0.4339 & 153 (53.9\%) & 0.5693 & 69 (56.1\%) & 0.5659 & 1 (50.0\%) & 0.0554 \\ \hline
		  Hybrid DNR-GA & 235 (37.3\%) & 0.02 & 307 (51.3\%) & 0.0181 & 405 (54.7\%) & 0.0192 & 151 (53.2\%) & 0.0141 & 69 (56.1\%) & 0.0125 & 1 (50.0\%) & 0.0586 \\ \hline
    \end{tabular}
	\end{center}
\end{table*}

Tables \ref{tab:results_by_rating_iec} and \ref{tab:results_by_rating_nema} show the convergence and average squared error rates for the IEC and NEMA motor data sets respectively, with the data sets subdivided by rated power. It is observed that the convergence and average error rates are not uniformly distributed across the full range of motor rated powers. 

Of interest is the poor performance of all algorithms on smaller motors, particularly motors rated below 4kW, where convergence rates are between 0.3\% and 12.3\% for IEC motors and between 1.4\% and 37.3\% for NEMA motors. Performance begins to improve for all algorithms as the motor size is increased. For motors $\geq$90kW, the convergence rates of the hybrid algorithms improve to $>$60\% for the IEC motors and $>$50\% for the NEMA motors. It should be noted that dynamic modelling is least likely to be performed on individual small motors, since they are often aggregated as lumped loads in power system studies.


\section{Conclusion}
In this paper, a number of algorithms were investigated for the estimation of induction motor parameters based on manufacturer data. Simulations on a large data set of IEC and NEMA motors showed that the conventional Newton-Raphson algorithm performs quite poorly. Hybrid algorithms were introduced as an alternative to the conventional NR algorithm and computer simulations suggested that the proposed hybrid algorithms show promise as a parameter estimation tool, with large improvements in convergence and error rates over the conventional algorithms.

The key drawback for the hybrid algorithms is their computation time, which depending on the algorithm settings, can be significantly slower than conventional descent or genetic algorithms. In any case, it can be argued that motor parameter estimation for the purpose of power system studies is not particularly time critical and a slow computation time can be tolerated in return for better algorithm performance.


%

\appendices



\ifCLASSOPTIONcaptionsoff
  \newpage
\fi



%

%

\begin{IEEEbiographynophoto}{Julius Susanto}
(M'11) received the B.Eng and M.Sc degrees in electrical engineering from Curtin University, Perth, Australia, in 2001 and 2013 respectively. Since 2002 he has worked in industry for WorleyParsons Ltd, DIgSILENT GmbH, TransGrid and Synergy Engineering focusing on power system studies. He is a Chartered Professional Engineer in Australia.
\end{IEEEbiographynophoto}

\begin{IEEEbiographynophoto}{Syed Islam}
(M'83,  SM'93)  received  the  B.Sc.  from Bangladesh  University  of  Engineering  and  Technology, Bangladesh,  M.Sc.  and  PhD  degrees  from  King  Fahd University of Petroleum and Minerals, Saudi Arabia, all in electrical  power  engineering  in  1979,  1983 and 1988 respectively. 

He is currently the Chair Professor in Electrical Power Engineering at Curtin  University, Perth, Australia.  He received the IEEE T Burke Haye's Faculty Recognition  award in 2000. His research interests are in Condition Monitoring of Transformers, Wind Energy Conversion and Power Systems. He is regular reviewer for the IEEE Transactions on Energy Conversion, Power Systems and Power Delivery. Prof. Islam is an editor of the IEEE Transaction on Sustainable Energy. 
\end{IEEEbiographynophoto}





\end{document}